# From Nucleobases to DNA: Clustering-Triggered Emission and Pressure-Induced Emission Enhancement

Yijing Cui, Yu Song Cai, Xuchen Wang, Xiang Chen, Junhao Duan, Guangxin Yang, Zhipeng Zhao, Yuhao Zhai, Guanjun Xiao*, Bo Zou, and Wang Zhang Yuan*

**ABSTRACT:** The photophysical properties of deoxyribonucleic acid (DNA) are fundamental to life sciences and biophotonics. While previous studies have generally been restricted to fluorescence, attributing it to $\pi$-$\pi^*$ transitions and charge transfer within nucleobases in dilute solution, these understandings fail to explain the pronounced visible emission in physiological and aggregated states, and moreover, ignore the possible phosphorescence. Addressing this critical gap, we systematically investigate native DNA across its structural hierarchy, from nucleobases to single-stranded chains, under varying states. We demonstrate that DNA exhibits excitation-dependent emission in aggregates and moreover room-temperature phosphorescence (RTP) in the solid state. These behaviors are rationalized by the clustering-triggered emission (CTE) mechanism, where nucleobases and electron-rich nonaromatic moieties like sugar and phosphate synergistically contribute to DNA photophysics. High-pressure experiments reveal a 207-fold luminescence enhancement for nucleotides at 26 GPa, largely retained after decompression, underscoring the precise control of emission by intermolecular interactions. This study not only elucidates the intrinsic luminescence mechanism of DNA and but also establishes pressure modulation as a versatile approach for developing new nucleic acid-inspired luminescent materials.

## INTRODUCTION

As the genetic blueprint of life, deoxyribonucleic acid (DNA) is fundamental to the storage and transmission of genetic information and is central to virtually all biological processes.[1] While its structural and biochemical roles are well-established, understanding the DNA's intrinsic photophysical behavior is increasingly vital for both elucidating in vivo molecular events and advancing biomedical photonic technologies.[2] Traditionally, DNA has been considered virtually nonemissive in dilute aqueous solutions (*e.g.*, 10$^{-5}$ M), exhibiting extremely low fluorescence quantum yields ($\Phi$) below 10$^{-4}$ (Figure 1a), thus requiring external fluorescent dyes for imaging and detection.[3] However, recent findings challenge this view, demonstrating that DNA can exhibit appreciable fluorescence at physiological concentrations, such as within interphase nuclei or metaphase chromosomes.[4] And moreover, increasing results show the important role of electron-rich moieties, including amine ($NH_2$), carbonyl (C=O), and heteroatoms (*e.g.*, N, O, P), to the solid photoluminescence (PL) of biomolecules,[5-7] according to the clustering-triggered emission (CTE) mechanism.[5a,7] This growing evidence unveils a more complex photophysical landscape of DNA than previously recognized, highlighting the critical need to uncover the mechanisms underlying its intrinsic luminescence.

Traditional explanations for DNA luminescence, primarily based on $\pi$-$\pi$* transitions and charge transfer (CT) processes within aromatic nucleobases,[2a,8] fail to fully account for the unique emission behavior observed in aggregates, particularly, ignore the possible room temperature phosphorescence (RTP) emission. Moreover, despite their prevalence and significant potential, the photophysical contributions of electron-rich moieties have been largely underestimated. Herein, to gain further insights into DNA luminescence, we systematically investigate DNA luminescence across its structural hierarchy, spanning nucleobases, nucleosides, nucleotides, and single-stranded DNA. It is found that from nucleobases to native DNA, they all exhibit aggregation-induced emission (AIE) features, and moreover excitation-dependent PL and RTP in the solid state. These phenomena are understandable in terms of the CTE mechanism for nonconventional luminophores,[5a,7] indicating synergistic contributions from both nucleobases and nonaromatic moieties to DNA luminescence. Specifically, high-pressure results on ctDNA, ssDNA, cytosine (C), deoxycytidine (dC), and deoxycytidine 5′-monophosphate (dCMP) (Figure 1b) revealed a substantial PL enhancement in dC and dCMP solids. Upon compression up to 26 GPa, nucleotides exhibited a remarkable 207-fold PL enhancement (Figure 1b), which was largely retained after pressure release. This pressure-induced emission enhancement (PIEE) suggests the formation of stable emissive clusters, a phenomenon related to the CTE mechanism.[9,10] This study elucidates the mechanistic basis of DNA luminescence and demonstrates the efficacy of pressure modulation as a strategy for designing nucleic acid-based luminescent materials, with potential applications in biosensing, imaging, and optoelectronic devices.

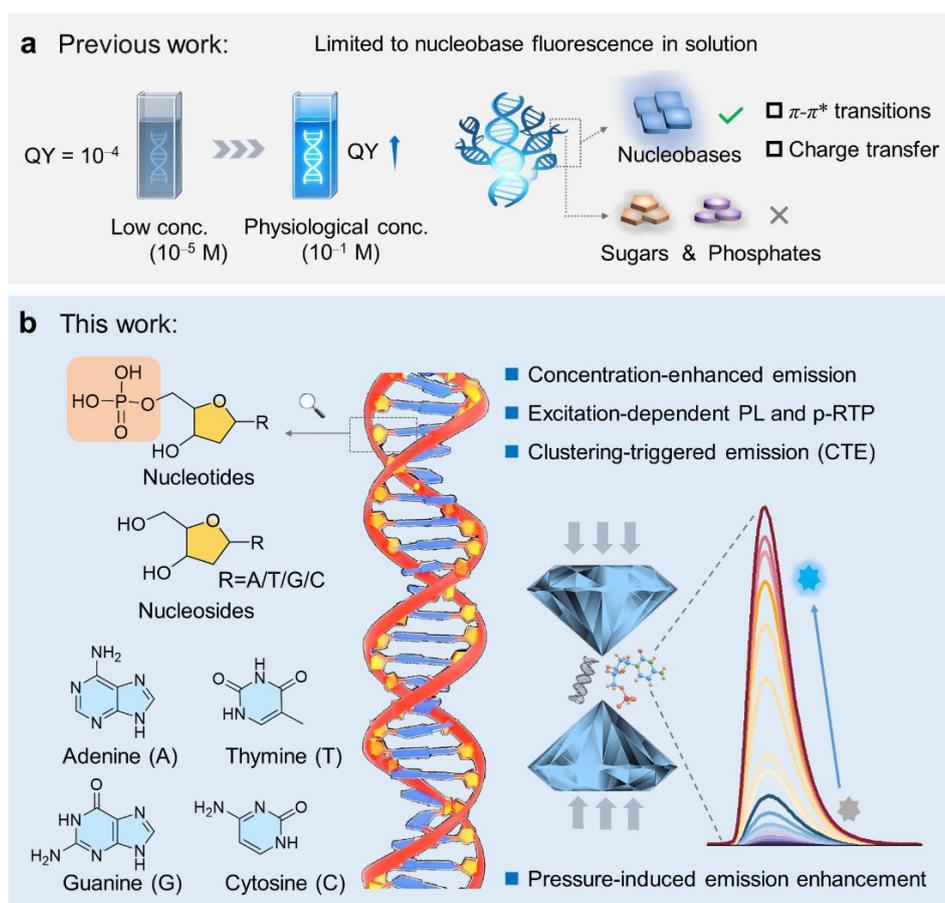

**Figure 1.** (a) Previous research on DNA luminescence and its underlying mechanism. (b) Structures studied herein and schematic illustration of PIEE phenomenon of dC.

## RESULTS AND DISCUSSION

We first investigated the PL of native DNA in aggregated states, employing calf thymus DNA (ctDNA) and salmon sperm DNA (ssDNA) as model systems.[11,12] As shown in Figures S1 and S2, both ctDNA and ssDNA exhibit pronounced concentration-enhanced emission in aqueous media. With increasing DNA concentration, fluorescence intensity increases significantly. While their dilute solutions are almost nonemissive, at a concentration of 20 mg mL$^{-1}$, the $\Phi$ reaches 1.6% for ctDNA and 1.5% for ssDNA, respectively (Figures S1 and S2), displaying characteristic AIE behaviors. Furthermore, both samples (10 mg mL$^{-1}$) displayed clear excitation wavelength ($\lambda_{ex}$)-dependent PL, with emission maximum ($\lambda_{em}$) shifting from approximately 400 to 520 nm as $\lambda_{ex}$ increased from 312 to 420 nm (Figures S1d and S2d). This spectral behavior closely resembles the intrinsic luminescence characteristics previously reported for natural proteins.[5b] These observations support the CTE mechanism as a rational framework for explaining the intrinsic luminescence of DNA, wherein electron-rich structural motifs (particularly deoxyribose and phosphate groups) undergo intermolecular interactions to form ordered emissive clusters. This clustering enhances molecular rigidity, promotes TSC, and ultimately facilitates radiative transitions.[5-7]

Meanwhile, ctDNA and ssDNA powders exhibit pronounced $\lambda_{ex}$-dependent emission (Figures 2a), with PL colors shifting from blue to yellow-green and CIE coordinates evolving from (0.21, 0.21) to (0.29, 0.43) for ctDNA, and from (0.19, 0.17) to (0.28, 0.44) for ssDNA (Figure 2b and S3a). Notably, for both samples the $\lambda_{em}$ shifts from ~400 to ~500 nm (Figure 2c and S3b), indicating the presence of diversified emissive clusters. Time-resolved measurements demonstrate that both samples exhibit nanosecond-scale multi-exponential decay kinetics upon excitation at 312 and 365 nm (Figure S4), indicating the coexistence of multiple radiative transition pathways. Remarkably, with a delay time ($t_d$) of 1 ms, both solids exhibit delayed peaks, shifting from 520 to 600 nm for ctDNA and 510 to 580 nm for ssDNA (Figures 2d and S3c), which are assignable to persistent RTP (p-RTP). These p-RTP emissions likely originate from the presence and clustering of n-electron-rich groups including nucleobases, carbonyl, phosphate, and hydroxyl, which enhance spin–orbit coupling (SOC) and facilitate intersystem crossing (ISC) of the clusters. At room temperature, the $\Phi$ of ctDNA and ssDNA solids are 7.2% and 6.6%, respectively, with corresponding phosphorescence quantum yields ($\Phi_p$) of 2.1% and 1.9%, indicating their dual capability in efficient fluorescence and appreciable phosphorescence (Figure 2e). To further suppress nonradiative decay pathways of triplet excitons and enhance luminescent performance, low-temperature measurements were performed. Upon cooling to 77 K, the $\Phi_p$ increased to 12.3% for ctDNA and 10.2% for ssDNA, accompanied by a significant extension in phosphorescence lifetime ($\tau_p$) (Figures 2e,f and S5). These results confirm that intermolecular interactions and conformational rigidification stabilize triplet excitons and enable remarkable RTP in DNA solids. The $\lambda_{ex}$-dependent PL and RTP of native DNA should originate from the clustering of aromatic nucleobases and nonaromatic electron-rich motifs like deoxyribose and phosphate groups.[13]

The vast diversity of DNA structures renders a comprehensive investigation of their photophysical properties challenging.[14-15] To gain deeper insights into the underlying luminescence mechanisms, C, dC, dCMP, and a single-stranded decamer of C (dC$_{10}$) were selected as representative model systems for systematic analysis of their luminescent behaviors (Figures 3a and S6–S15). Under excitation from 312 to 420 nm, C solids exhibit narrowband emission confined to the sky-blue region with a PL red-shift by 60 nm, alongside modest CIE coordinate changes and a narrow full-width at half-maximum (FWHM) of ~60 nm (Figures 3b and S16a,S17). This constrained behavior contrasts sharply with the broader, excitation-tunable emission of native DNA (Figure 2). Notably, incorporating a deoxyribose unit in dC markedly alters its photophysics, producing pronounced excitation dependence with color evolution from deep blue to green, a 90 nm red-shift, and a broadened FWHM of ~100 nm (Figures 3b,c and S16b,S18). Both C and dC solids display RTP, with red-shifted delayed peaks and millisecond lifetimes, while the $\Phi$ of dC reaches 6.1%, 1.69 times higher than that of C (Figure 3d,e and S17c). These results suggest that the introduction of deoxyribose in dC facilitates the clustering of electron-rich units, enhancing PL intensity and generating more diverse emissive species that differently response to varying excitations.[5a,d]

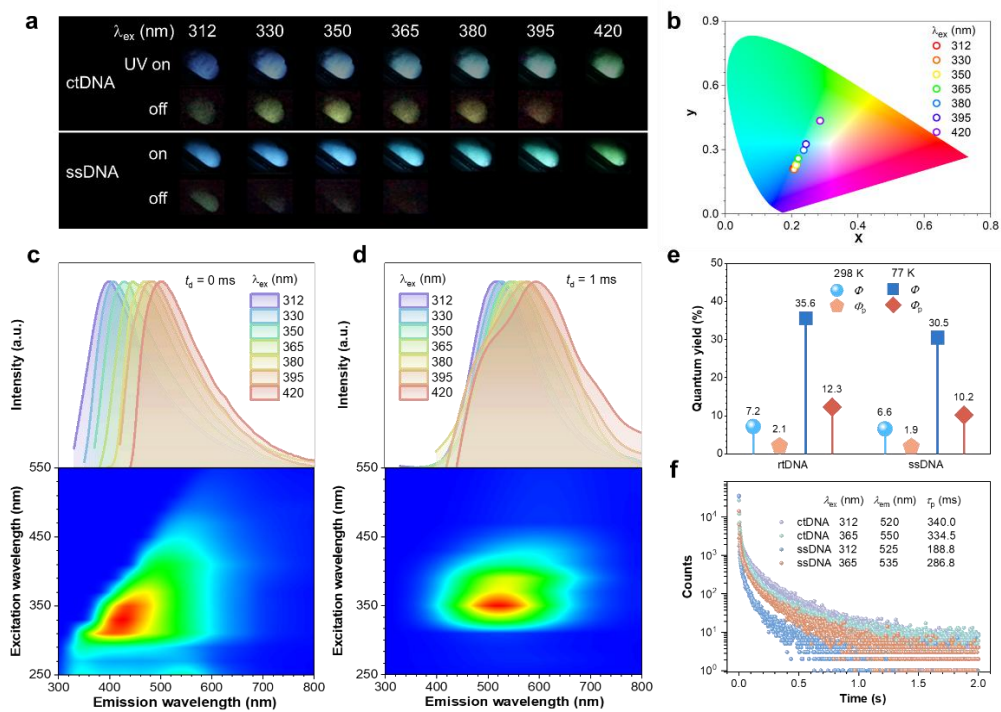

**Figure 2.** (a) Luminescent photographs of ctDNA and ssDNA solids. (b) CIE coordinate diagram of ctDNA. (c) Prompt and (d) delayed ($t_d$ = 1 ms) emission spectra (top) and corresponding emission-excitation mapping (bottom) of ctDNA solids. (e) $\Phi/\Phi_p$ and (f) $\tau_p$ of ctDNA and ssDNA solids. *Note*: Unless specified, all data were obtained at room temperature.

Unexpectedly, dCMP crystals, despite their structural similarity to dC crystals, exhibit a marked decrease in $\Phi$ (2.1%), although they still retain excitation-dependent luminescence (Figures 3e and S19). The efficiency of $dC_{10}$ solids further decreases to 1.8% (Figures 3e and S20). Visual inspection confirms a faint afterglow for dCMP solids, whereas no visible afterglow emission is noticed for $dC_{10}$. The $\tau_p$ and $\Phi$ values are also significantly reduced, suggesting that the introduction of the phosphate group and the elongation of the oligonucleotide chain may increase molecular flexibility and electrostatic repulsion, potentially disrupting the compact packing and rigid cluster conformation required for triplet stabilization. Furthermore, all aqueous solutions of three derivatives are nearly nonemissive below $10^{-3}$ M, however, above this threshold, dC and dCMP develop shoulder peaks and exhibit clear excitation-dependent emission at $10^{-2}$ M (Figure S21), featuring concentration-enhanced and $\lambda_{ex}$-dependent emissions. These PL behaviors of both solutions and crystals further confirm the rationality of the CTE mechanism.

Extending this approach to guanine (G) and its derivatives, dG and dGMP, reveals similar structure-dependent behaviors, exhibiting excitation-dependent PL and p-RTP with varying responses in the solids (Figures S22–S24). These universal PL and similar emission behaviors in nucleobases and derivatives can also be understood in terms of the CTE mechanism, considering the abundant of hetero-atoms and effective TSC in the solids.[16]

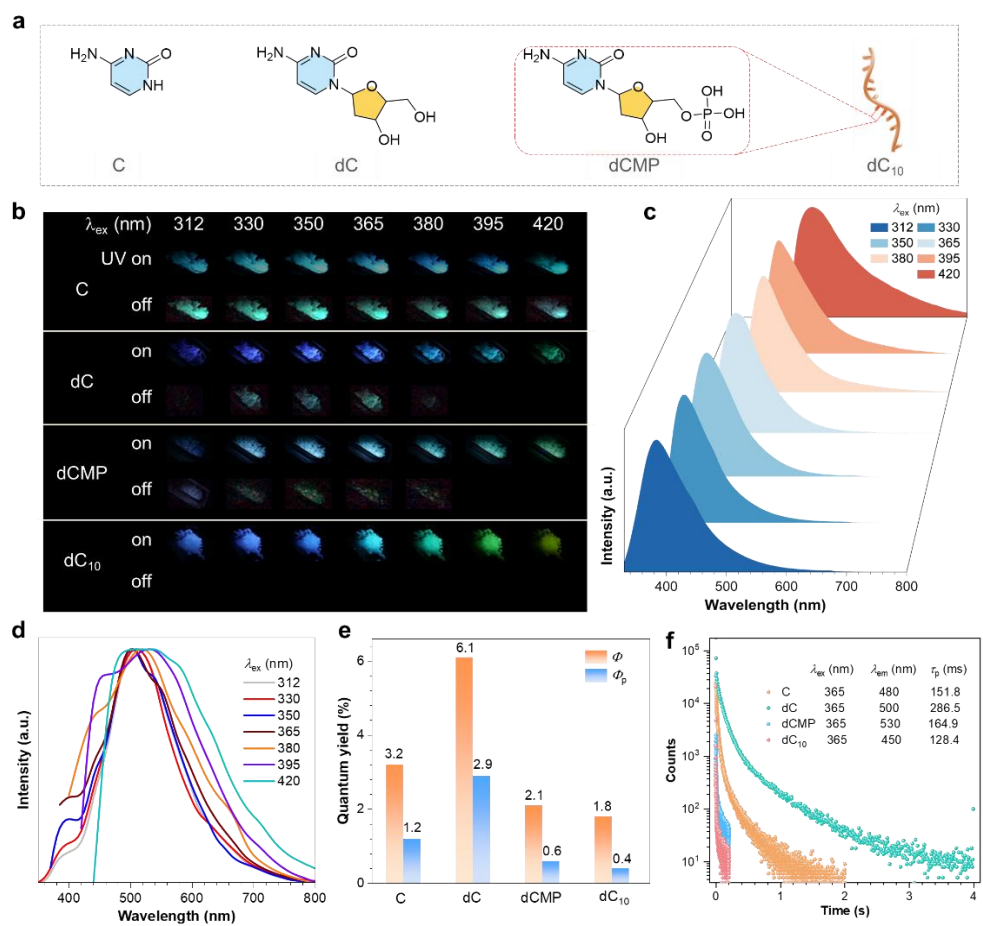

**Figure 3.** (a) Chemical structure or schematic diagram and (b) luminescent photographs of C, dC, dCMP, and dC$_{10}$ solids. (c) Normalized prompt and (d) delayed ($t_d$ = 1 ms) PL spectra of dC solids at different $\lambda_{ex}$s. (e) $\Phi/\Phi_p$ and (f) $\tau_p$ of C, dC, dCMP, and dC$_{10}$ solids under excition of 365 nm UV light.

To explore further insights into the structure-property relationships, we cultured single crystals of C, dC, and dCMP. As illustrated in Figure 4a, C packs in a $\pi$-$\pi$ stacked lamellar lattice with an interplanar distance of 3.324 Å, evidencing strong $\pi$-$\pi$ interactions. Introducing a deoxyribose disrupts continuous stacking, consequently, dC forms dense, directional hydrogen-bond networks (*e.g.*, C=O···H–O and C–O···H–N) and intra/intermolecular interactions amongst electron-rich units (*e.g.*, C–N···O), thus providing a structural basis for efficient TSC. In dCMP, the phosphate group induces a pronounced U-shaped distortion, increasing the base–base distance by about 0.06 Å. While hydrogen-bond and ionic (ion–dipole) interactions (*e.g.*, P–O···H–N, O–H···O–P, and N–H···O–H) are present, the anionic phosphate increases conformational freedom and electrostatic repulsion, leading to less directional, more isotropic packing compared to dC (Figure 4a).

Based on the single crystal structure and optimization, the intrinsic conformational rigidity of C, dC, and dCMP was assessed by single-molecule reorganization energy ($\lambda$), whose values are 11739, 6745, and 2291394 cm$^{-1}$ (Figure 4b), respectively. These results indicate that dCMP is the least intrinsically rigid (largest $\lambda$), whereas dC is the most intrinsically rigid (smallest $\lambda$). Dihedral contributions increase in the order C < dC < dCMP, reflecting a growing share of torsional reorganization and thereby reduced conformational rigidification. As a metric, λ quantifies the geometric alterations upon photoexcitation and reflects the contribution of intramolecular motions to nonradiative decay processes.[17] Because TSC is predominantly enhanced in the aggregated state, we examined the aggregates. Tetramer RMSD analysis further corroborates the rigidity hierarchy, showing dCMP > dC and minimal distortion for C (Figure S25). Hirshfeld surface (HS) analyses and the corresponding fingerprints revealed complementary trends in contact environments.[18] While the contact topology evolved from C's packing dominated by H–H interactions, dC characterized by expanded HS area (247.75 Å$^2$) and volume (262.95 Å$^3$), and a marked increase in short, directional O–H and H–H contacts (S26–S29). This dense polar network can stabilize clustered arrangements and facilitate TSC. In contrast, dCMP exhibited a packing motif with weaker directionality. While polar O–H and H–O contacts persisted, they are less directional, and the decrease in prominent N–C/C–N contacts, compared to dC, suggest limited contribution to electronic delocalization (Figures S28,S29).

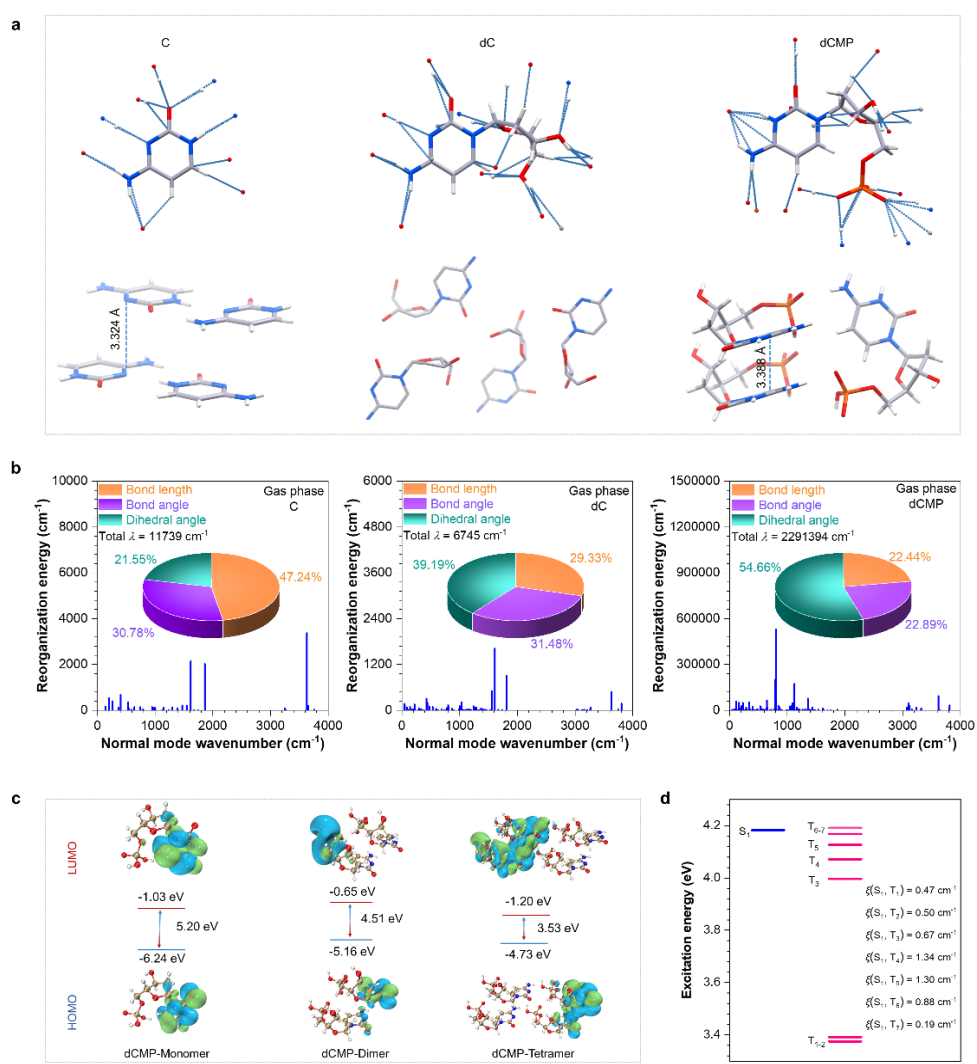

**Figure 4.** (a) Single-crystal structure with denoted intermolecular interactions and fragmental molecular packing of C, dC, and dCMP. (b) Plots of reorganization energy (λ) versus normal mode wavenumber of C, dC, and dCMP in the gas phase. Insets: proportions of bond length, bond angle, and dihedral angle contributed to the total λ. (c) HOMO and LUMO electron densities and energy levels for the monomer, dimer, and tetramer of dCMP. (d) The calculated excitation energies, SOC values (ξ) for dCMP dimer.

To gain further understanding, we examined excited-state properties by time-dependent density functional theory (TD-DFT) calculations. In dCMP, the phosphate group's participation at the monomer level lowers the LUMO to −1.03 eV and results in a wider HOMO–LUMO energy gap (5.20 eV) than in C (5.01 eV) or dC (4.84 eV) . Upon aggregation from monomer to dimer and tetramer, the frontier orbitals delocalize across the base–sugar–phosphate framework, narrowing the gap to 3.53 eV and evidencing strong TSC (Figures 4c and S30,S31). Noncovalent interactions (NCI) analysis further highlights this enriched interaction network by revealing abundant through-space interactions between adjacent dCMP molecules, a key factor for TSC (Figure S32). Meanwhile, electrostatic-potential (ESP) maps highlight distinct charge distributions, showing charge redistribution in the sugar ring and the phosphate group relative to C (Figure S33). Complementing these findings, SOC analysis indicates that the dCMP dimer exhibits $\xi(S_1,T_n)$ values up to ~1.34 cm$^{-1}$, consistent with efficient $S_1 \rightarrow T_n$ ISC and the feasibility of phosphorescence in stacked dCMP aggregates (Figure 4d). Notably, SOC values for C and dC dimers are also significant, suggesting that phosphorescence is a plausible pathway in these systems as well (Figures S34).

Further cryotemperature measurements strongly validated our theoretical conclusions regarding the impact of aggregation and structural rigidity on phosphorescence. Under cryogenic conditions, $\Phi_p$ for C, dC, dCMP and dC$_{10}$ solids exhibited substantial enhancements, increasing by 6.3-, 4.4-, 14.3-, and 14.5-fold, respectively, compared to ambient conditions. Corresponding $\tau_p$ reached 547.8, 1328.7, 296.2, and 191.6 ms (Figures S35–S37). These dramatic improvements underscore the crucial role of strengthened intermolecular interactions and enhanced conformational rigidity in stabilizing excited states, thereby facilitating efficient phosphorescence.

Given that pressure can significantly impact the distance and intra/intermolecular interactions among electron-rich moieties, we explored the PL performance of DNA solids under compression using a diamond anvil cell. For ctDNA, whose double helix is stabilized by hydrogen bonding and base stacking, a biphasic PL response was observed: a slight decrease betwen ambient pressure and 1.1 GPa, followed by a 1.3-fold increase at 15.8 GPa (Figure S38a,b). Upon decompression, PL partially declined, indicating incomplete recovery (Figures S38c,d).[10] In contrast, ssDNA, with its single-stranded nature and higher flexibility, exhibited red-shifted and diminished emission upon compression to 16.2 GPa (Figure 5a), but its PL rebounded by 1.31-fold after pressure release (Figures S39). These contrasting responses strongly suggest competive impacts on DNA luminescence under pressure, likely driven by the differential influence of compression on the packing and flexibility of double-stranded versus single-stranded structures, which in turn affect TSC and exciton stabilization.

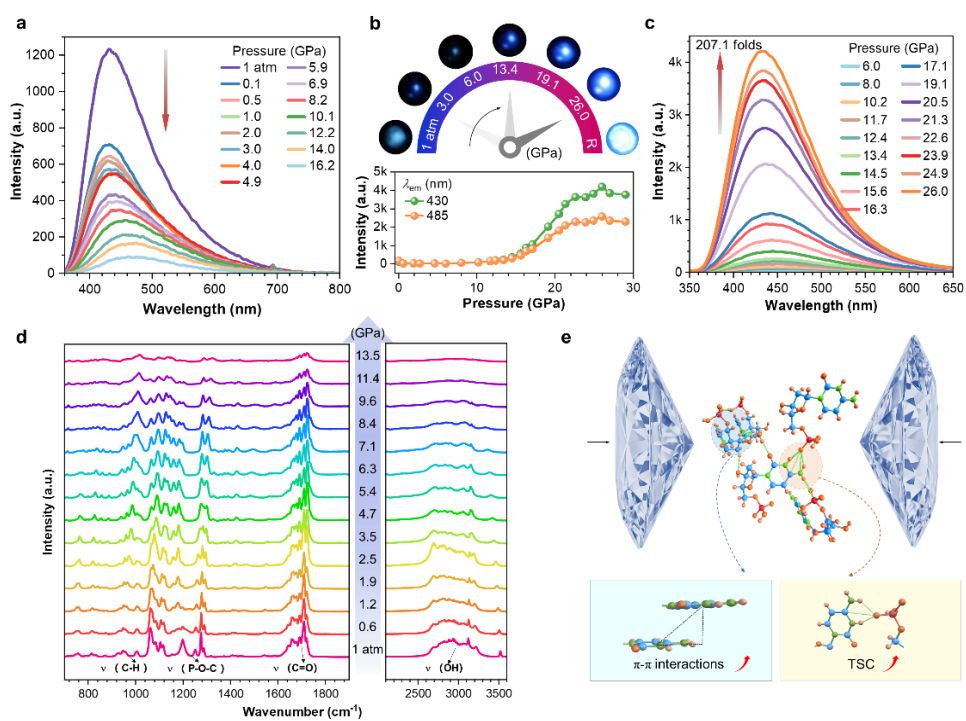

**Figure 5.** (a) PL spectra of ssDNA recorded at 1 atm–16.2 GPa. (b) PL peak intensity and corresponding ratio of the dCMP crystals during compression. (c) PL spectra of the dCMP crystals during compression under 6.0-26.0 GPa. (d) IR spectra of dCMP crystals upon compression. (e) Schematic diagram of the strengthening of π-π interactions and TSC.

Notably, C shows a continuous decrease in PL up to 15.2 GPa, with the intensity remaining low even after pressure release (Figure S40). In contrast, dC displays a complex response: an initial decrease up to 3.94 GPa, followed by a 4.1-fold increase with a 30 nm red shift upon further compression to 12.04 GPa, and a subsequent decline at higher pressures (Figure S41). A 32.9-fold enhancement is retained upon decompression (Figure S41f). Similary, dCMP initially shows a slight decrease up to ~6 GPa (Figure S42a), followed by a remarkable 207.1-fold increase and a 35 nm blue shift by 26 GPa (Figures 5b,c). This significant enhancement, attributed to denser packing and suppressed nonradiative decay, is followed by a slight intensity decline at even higher pressures (Figure S42b). Upon release, dCMP maintains a substantial 68.6-fold enhancement (Figure S42c), suggesting that pressure-induced packing reorganization preserves cooperative short-contact networks, thereby strengthening TSC.

To probe the structue changes, in situ high-pressure infrared (IR) measurement was conducted. As can be seen in Figure 5d, the P–O–C stretching band near 1 200 cm$^{-1}$ exhibits an initial blue shift up to ~6.3 GPa, followed by a red shift, reflecting phosphate backbone densification and enhanced spatial proximity. Simultaneously, P=O (~1 300 cm$^{-1}$) and C=O (~1 700 cm$^{-1}$) bands shift to lower wavenumbers, and O–H envelope (2 500–3 000 cm$^{-1}$) broadens/levels, consistent with strengthened H-bonding.[19] Notably, the out-of-plane C–H bending band (950–1000 cm$^{-1}$) [20] evolves from a single feature to a doublet above ~6.3 Gpa, with both components subsequently shifting to higher wavenumbers, indicating enhanced intermolecular coupling and altered base stacking. Collectively, these IR results reveal how high pressure tunes spatial proximity, π-π interctions, and TSC, thereby modulating dCMP PL.[21] Moderate compression, leading to phosphate backbone densification and strengthened base stacking, correlates with an initial PL reduction. However, further compression drives aggregate reorganization and enhanced electronic interactions among nonconvention luminophores, forming emissive clusters with enhanced conformational rigidity and boosted PL (Figure 5e). These findings robustly validate the CTE mechanism, demonstrating that PIEE offers a versatile tool for engineering luminescent DNA systems.

## CONCLUSION

In summary, this study presents a comprehensive investigation into the photophysical behavior of DNA and its constituent structural components. Our findings underscore the pivotal synergistic role of nucleobases and nonaromatic moieties acting in modulating DNA luminescence. Through the construction of a hierarchical model, ranging from individual nucleobases to single-stranded chains, we newly establish that all solid-state forms of these materials exhibit excitation-dependent PL and, notably, RTP emission. This phenomenon is rationalized by the CTE mechanism. Specifically, the aggregation of electron-rich moieties, particularly nonconventional chromophores (*e.g.*, $NH_2$, C=O, sugar, phosphate) leads to the formation of diverse emissive species, concurrently with enhanced conformational rigidity leading to a significant improvement in PL efficiency. Furthermore, high-pressure experiments effectively modulated the PL through the regulation of intermolecular distance, electronic interactions and conformational rigidity. In particular, dCMP crystals exhibit a pronounced PIEE response, achieving a 207-fold increase in PL intensity and retaining a 68.6-fold enhancement even after decompression. This work advances the fundamental understanding of DNA luminescence by shifting the focus from solely aromatic nucleobases to encompass both nucleobases and electron-rich nonconventional chromophores. Consequently, it establishes a crucial foundation for the development of new luminescent materials utilizing DNA-derived building blocks.

## ASSOCIATED CONTENT

**Supporting Information**
Experimental details, characterizations, crystallographic information, and theoretiical calculation details.(PDF)

## AUTHOR INFORMATION

**Corresponding Author**
**Wang Zhang Yuan** – *School of Chemistry and Chemical Engineering, Frontiers Science Center for Transformative Molecules, Shanghai Jiao Tong University, 800 Dongchuan Road, Shanghai 200240, China,* orcid.org/0000-0001-6340-3497, Email: wzhyuan@sjtu.edu.cn
**Guanjun Xiao** – *State Key Laboratory of High Pressure and Superhard Materials, College of Physics, Jilin University 2699 Qianjing Road, Changchun 130012, China,* E-mail: xguanjun@jlu.edu.cn

**Authors**
**Yijing Cui** – *School of Chemistry and Chemical Engineering, Frontiers Science Center for Transformative Molecules, Shanghai Jiao Tong University, 800 Dongchuan Road, Shanghai 200240, China*
**Yu Song Cai** – *School of Chemistry and Chemical Engineering, Frontiers Science Center for Transformative Molecules, Shanghai Jiao Tong University, 800 Dongchuan Road, Shanghai 200240, China*
**Xuchen Wang** –*State Key Laboratory of High Pressure and Superhard Materials, College of Physics, Jilin University 2699 Qianjing Road, Changchun 130012, China*

**Junhao Duan** – *School of Chemistry and Chemical Engineering, Frontiers Science Center for Transformative Molecules, Shanghai Jiao Tong University, 800 Dongchuan Road, Shanghai 200240, China*

**Guangxin Yang** – *School of Chemistry and Chemical Engineering, Frontiers Science Center for Transformative Molecules, Shanghai Jiao Tong University, 800 Dongchuan Road, Shanghai 200240, China*

**Zhipeng Zhao** – *School of Chemistry and Chemical Engineering, Frontiers Science Center for Transformative Molecules, Shanghai Jiao Tong University, 800 Dongchuan Road, Shanghai 200240, China*

**Yuhao Zhai** – *School of Chemistry and Chemical Engineering, Frontiers Science Center for Transformative Molecules, Shanghai Jiao Tong University, 800 Dongchuan Road, Shanghai 200240, China*

**Bo Zou** – *State Key Laboratory of High Pressure and Superhard Materials, College of Physics, Jilin University 2699 Qianjing Road, Changchun 130012, China*

## CONFLICTS OF INTERE

The authors declare no competing financial interest.

## ACKNOWLEDGMENT

This work was financially supported by the National Natural Science Foundation of China (U22A20250 and 52473185).

# TOC Graphic

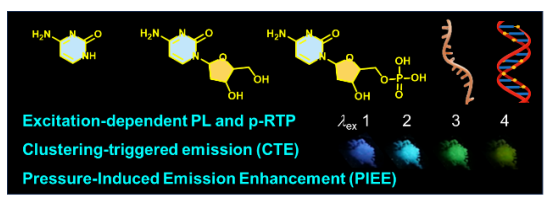